\documentclass[traditabstract]{aa} 

\usepackage{natbib} 
\bibpunct{(}{)}{;}{a}{}{,} 
\bibstyle{hapj}     
\usepackage{txfonts}
\usepackage{fixltx2e} 
\usepackage{graphicx} 
\usepackage{aas_macros}
\usepackage{subfig}
\usepackage{textcomp}

\def\widefield{wide-field}
\def\Widefield{Wide-field}
\def\apriori{\emph{a priori}}

\def\adhoc{\emph{ad hoc}}

\def\fig{Fig.}

\def\Fig{Figure}
\def\Figs{Figures}
\def\sect{Sect.}
\def\sects{Sects.}
\def\Sect{Section}

\def\tab{table}

\def\Tab{Table}

\def\eqn{equation}
\def\eqns{equations}

\def\etal{et~al.}

\def \rah   {\hbox{$^{\rm h}$}}
\def \ram   {\hbox{$^{\rm m}$}}
\def \ras   {\hbox{$^{\rm s}$}}
\def \fras  {\hbox{\hspace{0.05 em}$.\!\!^{\rm s}$}}
\title{VLBI imaging throughout the primary beam using accurate UV~shifting}
\titlerunning{VLBI Imaging throughout the primary beam}
 
\author{ J.S. Morgan\thanks{Currently at the International Centre for Radio Astronomy Research} \inst{1,2,3} \and
         F. Mantovani \inst{1} \and  
	 A.T. Deller  \inst{4} \and
	 W. Brisken   \inst{4} \and
         W. Alef      \inst{2} \and
	 E. Middelberg\inst{5} \and
	 M. Nanni     \inst{1} \and
         S.J. Tingay    \inst{3}
       }
\authorrunning{J.S. Morgan et al.}

\offprints{J.S. Morgan,\\
  \email{john.morgan@icrar.org}}

\institute{Istituto di Radioastronomia -- INAF, Via Gobetti 101,
 I-40129 Bologna, Italy
\and Max-Planck-Institut f\"ur Radioastronomie, Auf dem H\"ugel 69,
53121 Bonn, Germany
\and International Centre for Radio Astronomy Research, Curtin University,
GPO Box U1987, Perth, WA 6845, Australia
\and National Radio Astronomy Observatory, P.O. Box O, Socorro, NM 87801, USA
\and Astronomisches Institut, Ruhr-Universit\"at Bochum, Universit\"atsstr. 150, 44801 Bochum, Germany
}

\date{Received September 17, 2010; accepted December 13, 2010}

\abstract
{
For Very Long Baseline Interferometry (VLBI), the fringe spacing is extremely narrow compared to the field of view imposed by the primary beam of each element.
This means that an extremely large number of resolution units can potentially be imaged from a single observation.

We implement and test a technique for efficiently and accurately imaging large VLBI datasets.
The DiFX software correlator is used to generate a dataset with extremely high time and frequency resolution.
This large dataset is then transformed and averaged multiple times to generate many smaller datasets, each with a phase centre located at a different area of interest.

Results of an 8.4 GHz four-station VLBI observation of a field containing multiple sources are presented.  
Observations of the calibrator 3C345 were used for preliminary tests of accuracy of the shifting algorithm.
A high level of accuracy was achieved, making the method suitable even for the most demanding astrometric VLBI observations.
One target source (1320+299A) was detected and was used as a phase-reference calibrator in searching for further detections.
An image containing 13 billion pixels was constructed by independently imaging 782 visibility datasets covering the entire primary beam of the array.

Current implementations of this algorithm and possible future developments in VLBI data analysis are discussed.
}
\keywords 
	{
	Instrumentation: interferometers --
	Methods: data analysis --
	Quasars: individual: \object{1320+299} --
	Radio continuum: general --
	Techniques: interferometric --
	Techniques: image processing
	}

\begin{document}

\maketitle 

\section{Introduction}
For an interferometer, the fringe spacing or angular resolution $\theta$ is approximately $\lambda/D$ where $\lambda$ is the wavelength and $D$ is the baseline length.
The field of view is dictated by the solid angle of the sky to which the antenna is sensitive.
For parabolic antennas this can be approximated by a symmetric Gaussian with a half-power beam width $\Theta$ of the primary beam given by $1.2\lambda/d$ where $d$ is the diameter of the parabola.  
For VLBI, $D/d$ and therefore the number of resolution units across the primary beam $\Theta/\theta$ is particularly high (See \tab~\ref{table:arrays}).
For imaging purposes the number of pixels will be $\sim$3 times higher in both dimensions than the number of resolution units to ensure proper sampling of the dirty beam.
\newlength{\zerowidth}
\settowidth{\zerowidth}{0}
\begin{table}
	\begin{minipage}[t]{\columnwidth}
	\caption{
	Comparison of field of view in resolution units for different arrays.
	}
	\label{table:arrays}
	\centering
	\renewcommand{\footnoterule}{}  
	\setlength{\tabcolsep}{3.5pt}
	\begin{tabular}{l c r@{\,}l r@{\,}l}
		\hline\hline
Array				& $d$ (m)	& \multicolumn{2}{c}{$D$ (km)}	& \multicolumn{2}{c}{$D/d$}  \\
\hline
VLA\tablefootmark{a} 		&  25	&    &\hspace{\zerowidth}36	&   1&440\\ 
Merlin\tablefootmark{b} 	&  32	&    &                  217	&   8&680\\ 
EfMaMcNo\tablefootmark{c} 	& 100	&   1&334			&  10&334\\
EVN 				& 100	&  10&180			& 101&800\\
VLBA				&  25	&   8&611			& 344&440\\
\hline
	\end{tabular}
	\tablefoot{
	Diameter of largest antenna $d$;
	length of longest baseline $D$;
	the approximate number of resolution units across the primary beam $D/d\approx\Theta/\theta$.\\
	\tablefoottext{a}{A configuration.}
	\tablefoottext{b}{Without the Lovell 76-m.}
	\tablefoottext{c}{Array described in \sect~\ref{sec:observation}.}}
	\end{minipage}
\end{table}

This means that very large images can be generated. However there are three caveats:
(1) this technique is only useful if there are several sources sufficiently bright and compact for VLBI detection within the primary beam;
(2) it is necessary to correlate the data with sufficient resolution in time and frequency to avoid bandwidth and time-average smearing;
(3) there must be sufficient computing resources and appropriate techniques to handle the resulting large datasets.

Progress in all of these three areas has been made in the decade since post-processing computers have been able to handle the maximum output rates of hardware correlators.
\citet{Garrett:1999} review early efforts at \widefield\ imaging of objects including gravitational lenses and 
multiple supernova remnants located in the same galaxy \citep{Pedlar:1999}.

Hardware correlators generally place some constraint on the number of output channels and on the output integration time \citep[For example the JIVE Mark 4 correlator: see][]{Campbell:2004}. 
This in turn limits the possible field of view of the output dataset due to time and bandwidth smearing (see \sect~\ref{sec:amploss}).
Multiple sources can be imaged by using multiple passes through the correlator \citep{Rioja:1999} however this is infeasible for a large number of sources.

In spite of these limitations further progress was made including VLBI observations of the \textit{Hubble} deep field \citep{Garrett:2001} and the NOAO bootes field \citep{Garrett:2005}.
\citet{Lenc:2008} took data from low-frequency observations and analysed it using \widefield\ VLBI techniques (a mixture of multiple correlator passes and UV shifting).
This led to an unprecedented 28 detections.
Nonetheless, time and frequency resolution constraints necessitated the discarding of data from longer baselines with increasing distance from the phase centre; thus none of the above studies were able to image the entire area permitted by the field of view of the interferometer elements with full sensitivity and UV coverage.

Constraints on the time or frequency resolution are much less stringent in software correlators such as the DiFX software correlator \citep{Deller:2007} and are thus a good match to the requirements of \widefield\ imaging.
The first attempt to image objects across the entire primary beam without discarding data was made by \citet{Middelberg:2008}.

Technical developments in VLBI mean that bandwidths continue to increase.
The LBA currently works at a data rate of 512\,Mbit\,s$^{-1}$ with higher rates possible for some antennas \citep{Phillips:2009a}.
The EVN is pushing towards 4\,Gbit\,s$^{-1}$ from 1\,Gbit\,s$^{-1}$ \citep{Langvelde:2009}.
The ongoing VLBA sensitivity upgrade will increase the bandwidth from 512\,Mbit\,s$^{-1}$ to 2\,Gbit\,s$^{-1}$ and then to 4\,Gbit\,s$^{-1}$ \citep{Walker:2007}, increasing continuum sensitivity by a factor of approximately three.  
This will increase the density of detectable sources on the sky.

\section{Methods}
\subsection{Correlation and generation of a visibility dataset}
\label{derivation}
In VLBI correlation it is necessary to generate a delay model to align the incoming sampled data as if the antennas were all on a plane perpendicular to the point on the sky being observed \citep{Sovers:1998}. 
In order to achieve the accuracy required to ensure picosecond timing precision (required for microarcsecond precision astrometry) the model used is necessarily extremely complex.

The delay model dictates the `phase centre' of the data: the location where the residual fringe delay and rate are zero.
If the field only contains a single point source at the phase centre, the visibility phases will be static along the time and frequency axes (ignoring un-modelled instrumental and atmospheric effects).
A single source that is far from this phase centre will have visibilities that rotate rapidly in phase along the frequency and time axes due to the residual delay and residual delay rate respectively.

If the data are over-averaged this will lead to a reduction in flux density, and increasing uncertainty in the position of the source.
In an FX correlator, fixed blocks of samples are selected for each baseline and Fourier-transformed before correlation. 
Only samples in corresponding FFT blocks can be compared leading to further reduced sensitivity to sources distant from the phase centre \citep{Romney:1999}.

The baseline vectors $(u, v, w)$ are calculated along with delays.
These define the baseline with respect to the source for each time integration and baseline and are used when imaging to compute the residual visibility phase $\Delta \phi$ for sources away from the phase centre as
\begin{equation} 
	\Delta \phi = \frac{2 \pi}{\lambda} (l u + m v)
	\label{eqn:deltaphi}
\end{equation}
for a source displaced $(l, m)$ radians from the phase centre as a substitute for computing the full delay at that point.

Traditionally $u$ and $v$ have been calculated using a purely geometric model.
This neglects many effects, the most significant of which is differential aberration, a relativistic effect due to the orbital velocity of the Earth ($v_{\oplus}$) with a magnitude of $\sim v_{\oplus}/c$, up to about $10^{-4}$, implying image registration errors of $\sim 1$ pixel for shifts of $10^4$ pixels if not corrected.
This will lead to significant errors for \widefield\ imaging.

More accurate values of $u$ and $v$ can be derived using the delay model.
These are calculated by sampling the delay at two further points: one with a small offset in $l$, another with a small offset in $m$.
Along with the delay already calculated for the phase centre, the change in delay with respect to $l$ and $m$ can be calculated to give baseline vectors which make \eqn~(\ref{eqn:deltaphi}) accurate to first order and consistent with the delay model used:
\begin{equation} 
	(u, v, w) = c \left(\frac{\partial \tau}{\partial l} , \frac{\partial \tau}{\partial m} , \tau \right) .
	\label{eqn:betteruvw}
\end{equation}

An additional benefit of using \eqn~(\ref{eqn:betteruvw}) to calculate baseline vectors is that it provides a well-defined prescription even in the case of very near-field targets, such as planets, asteroids and spacecraft.
The DiFX software correlator produces data with improved baseline vectors as described above.

The output of the correlator is a dataset consisting of 
(1) the right ascension and declination of the phase centre; 
(2) baseline vectors for each baseline and time integration;
(3) the visibility data themselves.
These visibility data are 4-dimensional with baseline, time, frequency and polarisation axes.
Each datum is a complex number representing phase and amplitude.
Other data usually included are the delay model used and various data useful for calibration and imaging.

\subsection{\Widefield\ VLBI data reduction}
An obvious approach for imaging over the entire beam is to use the large visibility dataset directly \citep{Garrett:2001} at least for an initial search for detections \citep{Lenc:2008}. 
Data reduction algorithms such as `IMAGR' and `VTESS' in AIPS are able to account properly for time and bandwidth smearing, working on datasets with multiple channels, and even shifting the phase centre of the image.
Some efforts are being made to parallelise these algorithms \citep{Bhatnagar:2009} however no currently available software suitable for VLBI data reduction allows this as far as we are aware.
Even with the fastest CPUs currently available data processing time becomes infeasible for VLBI images larger than a small fraction of the primary beam.
Moreover, direct \widefield\ imaging imposes added complications such as correcting smearing due to non-coplanar effects which quickly lead to overwhelming computing requirements.
Finally as shown in \sect~\ref{derivation}, current datasets provide information on how the delay varies across the \widefield\ image only to first order. 
This will inevitably lead to errors in position for sources far from the phase centre unless a sophisticated delay model is used for correction at the imaging stage.

Provided that the baseband data has been recorded (not necessarily the case for e-VLBI) it is possible to correlate more than once using a different delay model each time.
This permits the generation of several datasets phase-centred on different areas of the primary beam.
However, unless the correlator is capable of correlating multiple phase centres simultaneously, the correlator time required will soon become prohibitive for a large number of widely-separated sources.

\subsection{Transforming visibility datasets}
\label{largeuv}
Another possibility is to correlate once with time and frequency resolution sufficient to avoid significant smearing on any part of the field that is to be imaged.
This large single dataset can then be transformed to be phase-centred on a different point. 
Since time-average and bandwidth smearing effects increase radially from the phase centre, the data can be averaged in time and frequency before writing to disk.
Similarly, non-coplanar effects also increase radially from the phase centre and so are mitigated.
This shifting can be done multiple times, in parallel if necessary, generating multiple datasets.
These can then be imaged separately.

\subsubsection{The transform}
In order to transform the data to a new phase centre we must:
(1) replace the phase centre coordinates;
(2) replace the baseline vectors with those calculated for the new
phase centre;
(3) add a phase shift to every visibility. 
The shift is time, frequency and baseline dependent but the same for all polarisations.

The dataset can then be averaged in time and frequency. 
The degree of averaging possible will depend on the image size desired. 
If all of the primary beam is to be imaged, the averaging will depend on the spacing of the phase centres.
If individual sources are being picked out, it will depend on the source structure and the uncertainty in the \apriori\ position of the source.
In any case uncertainties in the antenna delay calibrations typically limit averaging to $\sim$5 seconds in time and $\sim$0.5~MHz in frequency for most centimetre wavelength observations.

\subsubsection{Limitations of current software}
Interferometry data reduction packages commonly have algorithms for performing this shift, for example AIPS has the task `UVFIX'.
However these algorithms are designed for shorter shifts and positional accuracy is lost on longer shifts. 
This is demonstrated in \sect~\ref{verification} and has also been noted by others \citep[see][]{Lenc:2008,Middelberg:2008}.
The averaging must be performed separately from shifting and the data must be sorted on disk multiple times requiring more disk usage.
Moreover, extremely large datasets cannot be handled by AIPS efficiently, if at all\footnote{Specifically there is a limitation on $n_{IF} \times n_{channels} \times n_{polarisations}$ (currently 32768) when loading the dataset.}.
\subsubsection{Accurate UV shifting}
\label{sec:uvshifting}
By generating two delay models, one for the correlation phase centre and another for the desired phase centre we can transform the data very accurately.

Consider $n$ antennas located at different time-varying positions with respect a source at phase centre $\vec{s}$.
Each antenna receives and records a time-varying voltage $v_n(t)$.
During correlation the delay models $\tau_n(t)$ will be used to shift the functions $v_n(t)$ to $v_n(t+\tau_n(t))$ so that a wavefront emanating from $\vec{s}$ will be aligned for all $n$. 
We want to take a dataset correlated with a phase centre at $\vec{s'}$ (with delay models $\tau'_n(t)$) and transform it so that it is identical to a dataset correlated with phase centre $\vec{s}$.

Consider a baseline of antennas $i$ and $j$.
Points in $v_i(t)$ and $v_j(t)$ which would have been adjacent after the application of $\tau_n$ are separated by $\delta\tau$ after the application of $\tau'_i(t)$ and $\tau'_j(t)$. 
This separation $\delta\tau$ is the therefore the delay that we will have to apply in order to transform our data to the new phase centre.
To a first approximation it is given by $\delta\tau_0$ which is simply the difference between the two delay models for any time $t$:
\begin{equation}
	\delta\tau_0 = (\tau'_i(t)-\tau_i(t)) - (\tau'_j(t)-\tau_j(t))\, .
	\label{eqn:dt0}
\end{equation}

However this simplistic correction assumes that the delay models are not changing with time (i.e. $\dot{\tau}_n = \dot{\tau}'_n = 0$ where the dot notation is used for time derivatives).
Essentially, we have estimated the delay correction at time $t$, when in fact we should have estimated the correction for each antenna $n$ at time $t + (\tau'_n(t)-\tau_n(t))$ in order to take into account the change in the delay model in the interval $\delta\tau$.
Thus, approximating the delay model as a linear function over this short span of time, a better approximation of $\delta\tau$ can be made:
\begin{equation}
	\delta\tau_1 = (\tau'_i(t)-\tau_i(t))(1+\dot{\tau}'_i(t)) - (\tau'_j(t)-\tau_j(t))(1+\dot{\tau}'_j(t)) \, ,
	\label{eqn:dt1}
\end{equation}
where we have also assumed that $\dot{\tau}_n = \dot{\tau'}_n$.
For ground-based VLBI, $\delta\tau_0$ and $\delta\tau_1$ typically differ by approximately 1 part in $10^6$, while the difference between $\delta\tau_1$ and $\delta\tau$ is less than 1 part in $10^8$.
The latter is shown empirically in \sect~\ref{verification}.

It is quite common to refine the delay model in post-processing using data not available at the correlator at correlation time (examples include using improved Earth orientation parameters). 
All delay calibration should properly account for the time-variation of the delay model.
However in most cases the shifts in the delay model are small enough that \eqn~(\ref{eqn:dt0}) is sufficient.

Finally the delay must be applied to the (Fourier transformed) visibilities. 
The phase shift $\phi$ to be applied is simply 
\begin{equation}
	\phi = 2\pi(\nu\cdot\delta\tau\bmod{1})
	\label{eqn:phi}
\end{equation}
for each sky frequency $\nu$.

\subsection{Correcting for amplitude losses}
\label{sec:amploss}
Amplitude loss to due to time-average and bandwidth smearing were described as averaging losses in \sect~\ref{derivation} and elsewhere (see \citet{Bridle:1999}; \citet[Chap.~6.4]{Thompson:2001}).

The time-average smearing of each shifted visibility amplitude ($R_s$) compared to the amplitude of the same source at the phase centre ($R_0$) is given by
\begin{equation}
	\frac{R_s}{R_0} = \mathrm{sinc}\left(\frac{2\pi\cdot\dot{\tau}\cdot t_{int}}{2}\right)
	\label{eqn:timesmear}
\end{equation}
where $t_{int}$ is the pre-shift accumulation period and $\dot{\tau}$ is the time derivative of $\delta\tau$.

In an FX correlator such as DiFX where only a single fast Fourier transform (FFT) is performed on each sample, the triangular lag function \citep[Fig.~8.17]{Thompson:2001} results in a reduced visibility amplitude for an offset continuum source. 
This is given by
\begin{equation}
	\frac{R_s}{R_0} = 1 - \delta\tau\cdot\delta\nu
	\label{eqn:lagcorr}
\end{equation}
where $\delta\nu$ is the width of each spectral channel\footnote{Assuming the baseband data is sampled at the Nyquist rate and an FFT size in samples of twice the number of spectral channels}. The lag function entirely dictates bandwidth smearing effects for an FX correlator.

We can also approximate the bandwidth smearing by analogy with the time-average smearing case (i.e. assuming a rectangular bandpass for each spectral channel).
Assuming that $R_0 - R_s \ll 1$ and replacing $\dot{\tau}\cdot t_{int}$ in \eqn~(\ref{eqn:timesmear}) with $\delta\tau\cdot\delta\nu$ we get
\begin{equation}
	\frac{R_s}{R_0} \approx 1 - \frac{\left(\pi\cdot\delta\tau\cdot\delta\nu\right)^2}{6}\,.
	\label{eqn:bwcorr}
\end{equation}
This is far less than that predicted by \eqn~(\ref{eqn:lagcorr}) and makes a case for overlapping FFTs in an FX correlator.

For full accountability, both delay models should be stored for each visibility in the output dataset along with the pre-shifted integration time and spectral channel width and the FFT setup of the correlator.
This would allow full correction in post-processing.
In the results we present in \sects~\ref{3comp} and \ref{sec:allpb}, all amplitude losses have been kept sufficiently small to be ignored.

\subsection{Correcting for the primary beam response}
Imaging to the edge of the primary beam is done routinely with connected-element arrays.
In this case a correction is applied to the intensity across the final image.
However for heterogeneous arrays it is more appropriate to apply an antenna by antenna correction.
This is achievable using the multiple phase centre technique if we assume that the primary beam response is constant across each sub-image.
The primary beam power function of a \emph{baseline} is given by
\begin{equation}
	\label{eqn:baselinebeam}
	A_{12}(l, m) = F^*_1(l, m)\times F_2(l, m)
\end{equation}
where $F_1$ and $F_2$ are the voltage patterns of the two antennas in the $l, m$ plane \citep[this is discussed in the context of VLBI arrays in][]{Strom:2004}.

The primary beam corrected amplitude of each visibility $I_{ijr}$ on the baseline $i, j$ in a dataset phase-centred in the $l, m$ plane can be calculated from the observed amplitude $I_{obs}$:
\begin{equation}
	\label{eqn:baselinecorrection}
	I_{ij}(l, m) = \frac{I_{obs}}{A_{ij}(l, m)}
\end{equation}
where the amplitude of the power function has been scaled to 1 at the centre of the primary beam.

It is also possible to calculate the theoretical RMS noise limit:
\begin{equation}
	\Delta I_m = \frac{1}{\eta_s\cdot\sqrt{2\cdot\Delta\nu\cdot\tau_{int}}}\times \left(\sum_{i=1,\,j=i+1}^{i=N-1,\,j=N} \frac{|A_{i,j}(l, m)|^2}{SEFD_i\times SEFD_j}\right)^{-1/2}
	\label{eqn:noisearraywidefield}
\end{equation}

In \Sect~\ref{sec:allpb} a simple truncated sinc function is used to model the primary beam.
The limitations of this simple model should be noted.
First, the primary beam is different at each frequency.
Second, the primary beam correction may be different for each polarisation due to offset feeds.
This, and any other non-radial function for the voltage, will lead to a time-varying primary beam for antennas with an alt-azimuth mount.
A full treatment of these effects for the VLBA is discussed in \citet{Middelberg:2010}.

\section{Observations of the 1320+299 Complex}
With the aim of testing the performance of the UV shifting method described above, we organised a four-station \adhoc\ VLBI observation.
The target field selected was that of the quasar 1320+299 (\object{4C\,29.48}).
The data recorded at each station were processed with the DiFX software correlator installed at the Istituto di Radioastronomia (IRA), Bologna, Italy.
Below we describe the target source, the observation parameters and the initial correlation.

This source was first observed at 1.4\,GHz and 5\,GHz using the Westerbork Synthesis Radio Telescope \citep{Feretti:1982} and later with the VLA in A and B configurations at 1.4\,GHz, 5\,GHz and 15\,GHz \citep[][herein referred to as Cornwell \etal]{Cornwell:1986}. 
The source was also observed by VERA at 22\,GHz \citep{Petrov:2007} with the component B chosen as the phase centre, however no detection was reported.

\begin{figure}
\resizebox{\hsize}{!}{\includegraphics[trim=35 0 48 0, clip, angle=270]{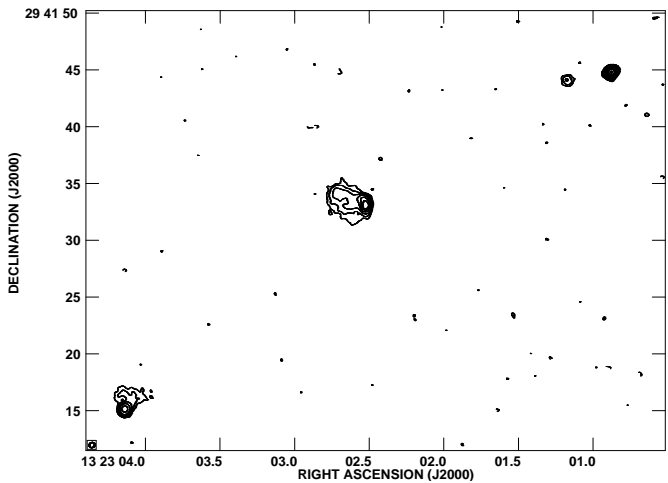}}
\caption{Image of the 1320+299 showing the three components. VLA in A configuration at 4.8\,GHz; synthesised beam shown at the bottom left; contour levels -0.7, 0.7, 1.4, 3, 8, 20, 40, 100, 200\,mJy beam$^{-1}$.}
\label{fig:abc}
\end{figure}

VLA and WSRT observations reveal three aligned components:
a 20th magnitude quasar is coincident with component A (see \fig~1, in Cornwell \etal).
\Fig~\ref{fig:abc}, which we have generated from data in the VLA archive shows the relative locations of the three components.

Component A has a one-sided morphology: a prominent radio core, unresolved by the highest-resolution observations carried out so far (Cornwell \etal\ VLA A configuration at 15\,GHz). It has peak brightness 210\,mJy beam$^{-1}$ at 5\,GHz and an integrated spectral index $\alpha$ of -0.42 (where S $\propto\nu^\alpha$).
A weak outer component is connected to the core region by a jet-like extension. 

Component B has an asymmetric structure with a `jet' which bends towards the north-east, extending to a distance of approximately 3\arcsec\ from the peak.
It has a peak brightness of 75\,mJy beam$^{-1}$ at 5\,GHz and is also unresolved by the VLA in A configuration.
The integrated spectral index of this component is -1.12.

The VLA A-configuration 5\,GHz image of component C reveals it to have a feature which extends in the north-west direction.
It has a peak brightness of 24\,mJy beam$^{-1}$ at 5\,GHz and an integrated spectral index of -0.92.

The three components are polarised and the degree of polarisation at maximum brightness ranges between 5 and 19\,\%. 
No significant depolarization or rotation measure is seen.

As a whole, the structure of 1320+299 looks rather peculiar.  
There is no conclusive evidence for the three components A, B and C to be physically related;
none of the images from prior observations show any bridge of emission connecting the components.
On the other hand, according to Cornwell \etal\ the probability of an unrelated source with $\lambda$20\,cm flux density between 250 and 450\, mJy lying within $\sim$25\arcsec\ of the central component B is only $\sim5\times10^{-5}$.
Cornwell \etal\ do not reach any firm conclusion about the classification of the components of the 1320+299 complex. 
Our observation, the first dedicated VLBI observation of 1320+299 to our knowledge, is also aimed at a better understanding of this puzzling complex.

\subsection{Observation} 
\label{sec:observation}
We observed the 1320+299 complex making use of an \adhoc\ network of antennas, namely Effelsberg (100-m), Medicina (32-m), Matera (20-m), and Wettzell (20-m).
The data were recorded using four base-band converters of the MarkIV acquisition system and MK5A disk recorder with a sample rate of $16\times10^6$ samples\,s$^{-1}$ and four-level sampling for all baseband channels.
The four observing frequencies were 8405.49, 8413.49, 8421.49, and 8429.49\,MHz.
Only the upper side band of 8\,MHz was recorded for each channel.
The observing session started on 12th December 2007 at 05:00:00 UT and stopped at 08:00:00 UT.
We were limited in our observing time by having only 1 Terabyte of storage available at the IRA computing centre at the time.

The observations were carried out using the phase-reference technique: switching between the target source 1320+299 and the calibrator with a duty cycle of 3 minutes on target and one minute on a point source phase calibrator.
The phase reference technique was required to increase our chances to detect fringes from the individual components A, B, and C and as a possible first epoch for future measurements of absolute component motions.
The VLBA calibrator \object{J1329+3154} (flux density 0.91\,Jy at 8.4 GHz), approximately 2.5\,\degr from the 1320+299 complex, was used as phase-reference source.

\subsection{Correlation with the DiFX software correlator at IRA}  
The MK5A disk packs recorded at each station were sent to the Medicina Radio Observatory, read back on a MK5A unit and loaded onto hard disks.
These were then transferred to IRA. 

The correlation of the data was carried out on a 12-node cluster.
Each node consists of two quad-core CPUs and the nodes are connected by 10\,Gigabit Ethernet.
Connected to the cluster is a RAID system served to the cluster using the GPFS file system.
Throughout the correlation the CPU and network utilisation was very low since the bottleneck was reading the data from the RAID.
With the modest observing parameters used for this experiment we were able to correlate in approximately real time.

\section{Analysis}
\subsection{Implementation of the shifting algorithm}
CALC 9.1 is typically used to generate delay models for DiFX.
The algorithm described in \sect~\ref{derivation} was implemented by modifying the DiFX FITS conversion tool to allow a second delay model to be read.
This could be used to apply any kind of modified delay model to a correlated dataset with high precision.

\subsection{Verification of the shifting algorithm}
\label{verification}
First it was necessary to test the accuracy of the UV shifting technique described in sect.~\ref{sec:uvshifting}.
The most meaningful measure is to compare a dataset correlated on phase centre $\vec{s}$ and shifted to phase centre $\vec{s'}$ with a dataset correlated directly with phase centre $\vec{s'}$.

To compare the two resulting datasets the method outlined in \citet{Tingay:2009} for comparing two correlators was used:
a single sub-band is extracted and the differences between the phases of the two datasets are measured and assessed for any systematic error that could degrade the astrometric precision even after calibration and averaging.
This analysis is carried out twice: first the two datasets are vector-averaged along the time axis before comparison, then the process is repeated with spectral averaging.
In each case the shifted dataset is subtracted from the reference dataset. The error is the mean value of this new dataset.

A scan of the source 3C345 was used for this test. 
Imaging this source showed it to be unresolved with a flux density of 5 Jy.
One minute of data were correlated twice: once on the known position of the source and once with a 1\arcmin\ offset added to both the R.A. and Dec coordinates (equivalent to $\sim$10\,000 turns of phase for each of the three baselines).
The second dataset was then shifted back, using both the AIPS task UVFIX and our algorithm.
Further tests were then done with shifts 1 and 2 orders of magnitude larger respectively.
The smallest of the three shifts is similar in magnitude to those required to obtain the results presented in \sect~\ref{sec:allpb}.
The largest is equivalent to shifting beyond the edge of the primary beam for 10\,m dishes on a 10\,000\,km baseline.
In each case the correlation parameters were chosen to keep time-average and bandwidth smearing to a level of approximately 5\% for the first two shifts and 30\% for the final shift\footnote{The actual correlation parameters were: 128, 1024 and 2048 channels per 8\,MHz sub-band and integration times of 0.4, 0.032 and 0.0064\,s for the three shifts.}.

All of the datasets were averaged post-shifting to a 2\,s integration time and 32 channels per sub-band.
The outer 8 channels and first and last time integrations were discarded before analysis.

\begin{figure*}
	\centering
	\resizebox{17cm}{!}
	{
		\subfloat[]{ 
			\label{fig:shifttimeuvfix}
			\resizebox{8cm}{!}{\includegraphics[clip]{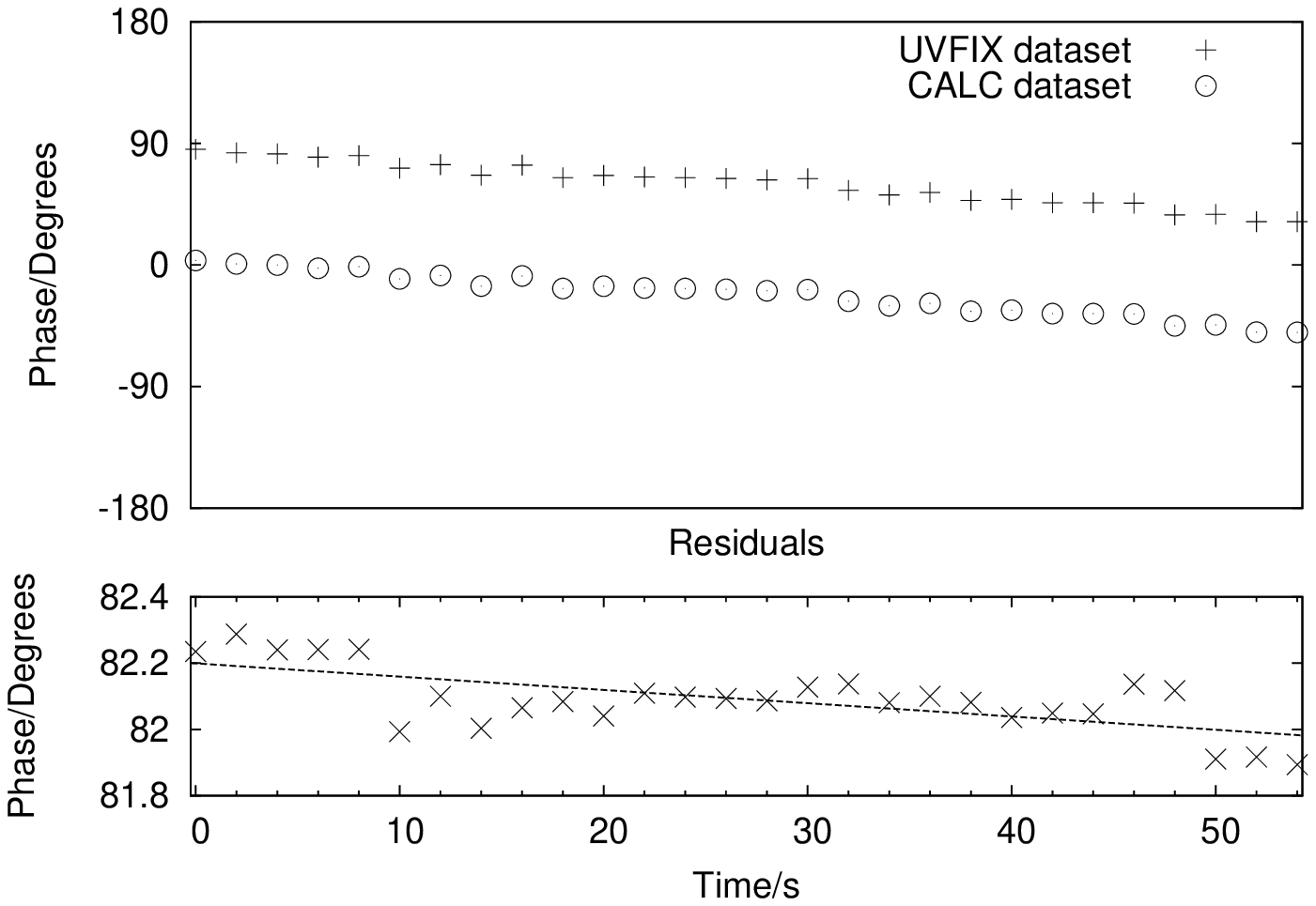}}}\quad
		\subfloat[]{
			\label{fig:shifttimecalc}
			\resizebox{8cm}{!}{\includegraphics[clip]{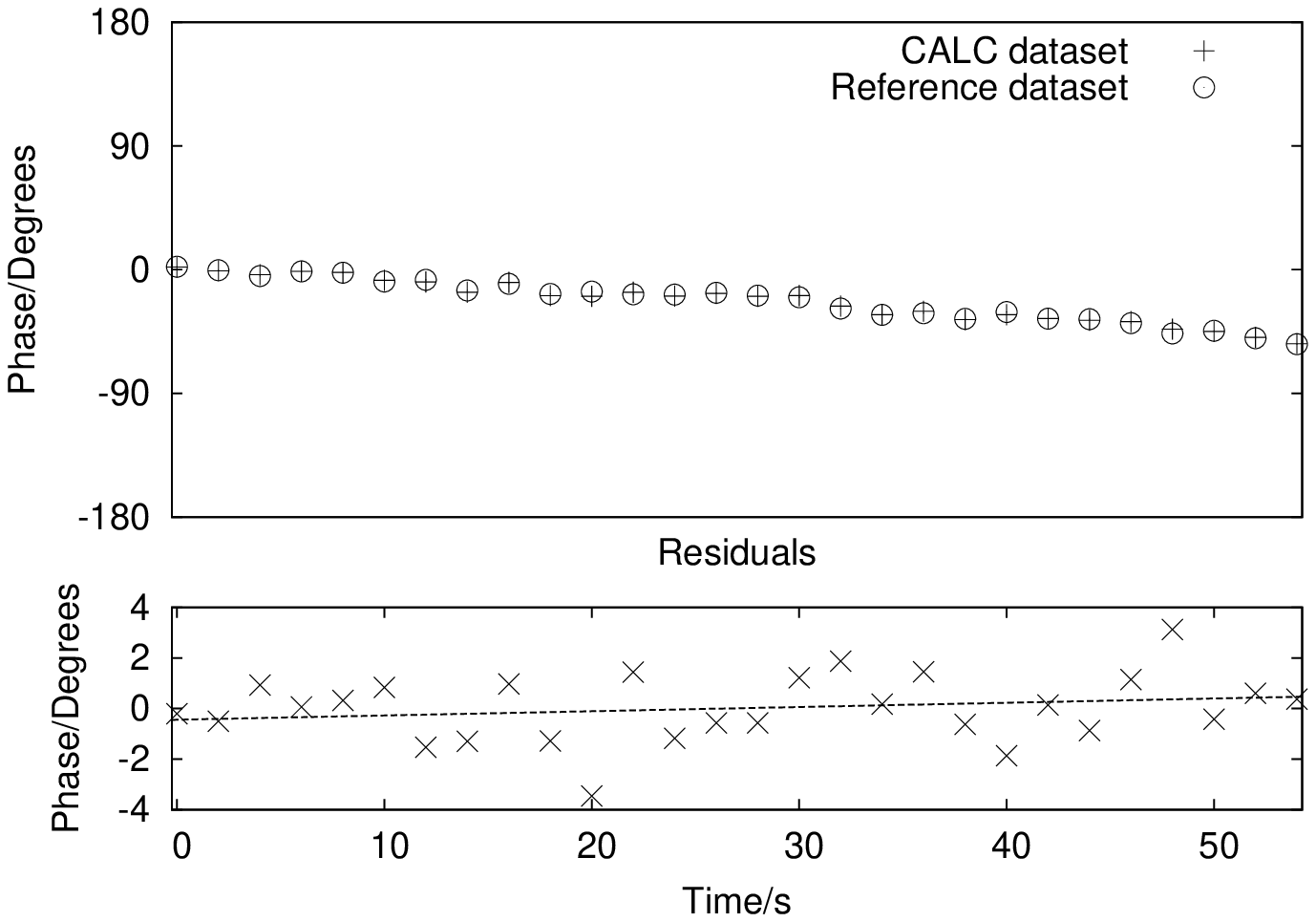}}}
	}
	\caption{Comparison of a dataset shifted using a second correlator delay model (our algorithm) with a UVFIX-shifted dataset (a) and with the reference dataset (b) for a single baseline (Matera-Medicina) with a positional shift of 1\arcmin\ N and 1\arcmin\ W. All spectral channels have been vector averaged. In (a) crosses  denote the UVFIX data and circles the dataset shifted with our algorithm. In (b) crosses denote the dataset shifted with our algorithm and circles the reference dataset.}
	\label{fig:shifttime}
\end{figure*}
\begin{figure*}
	\centering
	\resizebox{17cm}{!}
	{
		\subfloat[]{
			\label{fig:shiftchanuvfix}
			\resizebox{8cm}{!}{\includegraphics[clip]{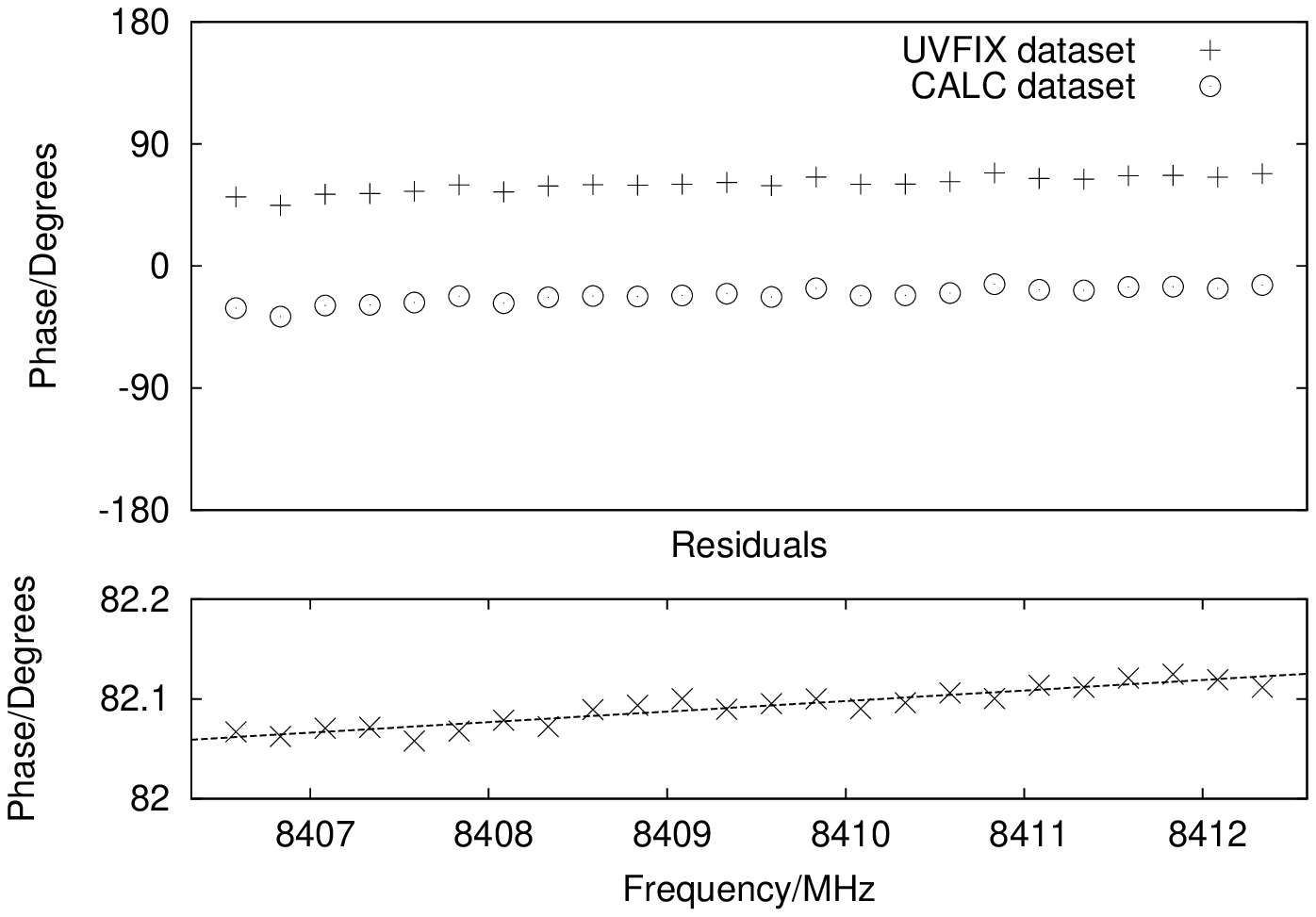}}}\quad
		\subfloat[]{
			\label{fig:shiftchancalc}
			\resizebox{8cm}{!}{\includegraphics[clip]{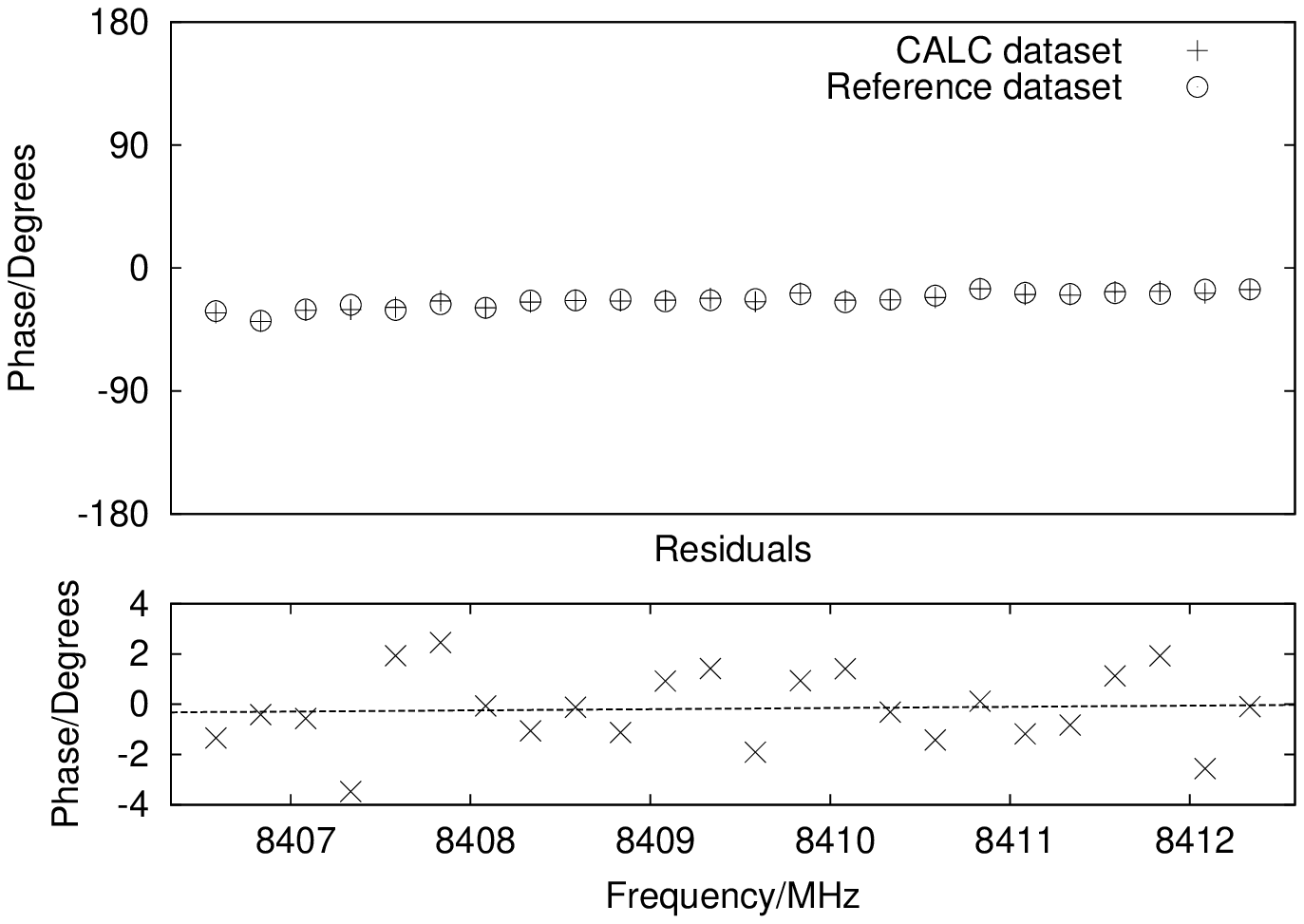}}
		}
	}
	\caption{Comparison of a dataset shifted using a second correlator delay model (our algorithm) with a UVFIX-shifted dataset (a) and with the reference dataset (b) for a single baseline (Matera-Medicina) with a positional shift of 1\arcmin\ N and 1\arcmin\ W. All time integrations have been vector averaged. In (a) crosses denote the UVFIX data and circles dataset shifted with our algorithm. In (b) crosses denote the dataset shifted with our algorithm and circles the reference dataset.}
	\label{fig:shiftchan}
\end{figure*}
\Figs~\ref{fig:shifttime} and \ref{fig:shiftchan} illustrate the results of these tests for a single baseline for the shorter shift.
The results for all shifts are summarised in \tab~\ref{tab:shiftresults}.
The $r^2$ value is given by
\begin{equation}
	r^2 = 1 - \frac{\sum_{i=1}^N\left(\delta\phi_i\right)^2}{\sum_{i=1}^N\left(\delta\phi_i-f_i\right)^2}
	\label{eqn:rs}
\end{equation}
where $\delta\phi$ is the residual phase error between the two datasets and $f$ is the expected value of $\delta\phi$ from the fit.

The low $r^2$ value for the linear regression suggests that the linear regression does not explain the variation of the points;  in other words, the residual error is mostly random.
Moreover there is no sign of any systematic error with increasing shift. 
The small random error is consistent with the expected noise introduced by smearing and would be an insignificant error even in the most precise VLBI measurements.
\begin{table*}
	\centering
	\caption{Accuracy of phase shifts.}
	\begin{tabular}{c r r r r r r r r r r r r} 
		\hline\hline
\multicolumn{1}{c}{Baseline} & 
\multicolumn{4}{c}{Shift 1} & 
\multicolumn{4}{c}{Shift 2} &
\multicolumn{4}{c}{Shift 3} \\
&
\multicolumn{1}{c}{Shift/turns} & 
\multicolumn{1}{c}{Error/\degr} & 
\multicolumn{1}{c}{$r^2_{freq}$} & 
\multicolumn{1}{c}{$r^2_{time}$} & 
\multicolumn{1}{c}{Shift/turns} & 
\multicolumn{1}{c}{Error/\degr} & 
\multicolumn{1}{c}{$r^2_{freq}$} & 
\multicolumn{1}{c}{$r^2_{time}$} & 
\multicolumn{1}{c}{Shift/turns} & 
\multicolumn{1}{c}{Error/\degr} &
\multicolumn{1}{c}{$r^2_{freq}$} &
\multicolumn{1}{c}{$r^2_{time}$} \\
\hline
MaMc & $5.5\times10^3$ & -0.18 & 0.017 & 0.087 & $5.5\times10^4$ & -0.21  & 0.034 & 0.077 & $5.4\times10^5$ &  1.90  & 0.46 & 0.408 \\
MaWz & $11\times10^3$  & 0.29 & 0.044 & 0.008 & $11\times10^4$  & -0.76  & 0.086 & 0.150 & $11\times10^5$  & -0.59 & 0.13 & 0.007 \\
McWz & $5.5\times10^3$ & -0.08 & 0.021 & 0.017 & $5.6\times10^4$ &  0.04  & 0.002 & 0.051 & $5.6\times10^5$ & -0.97  & 0.25 & 0.242 \\
\hline
	\end{tabular}
	\tablefoot{Shift magnitudes ($\nu\cdot\delta\tau$), errors and $r^2$ values. for 3 different shifts done using a second delay model. The $r^2$ is calculated from a least-squares fit to a linear model after vector averaging each spectral channel in time ($r^2_{freq}$) and after averaging all spectral channels in each time integration ($r^2_{time}$)}
	\label{tab:shiftresults}
\end{table*}
\begin{table*}
	\centering
	\caption{Accuracy of amplitude correction.}
	\begin{tabular}{c r r r r r r r r r r r r r r r} 
		\hline\hline
\multicolumn{1}{c}{Baseline} &
\multicolumn{4}{c}{Shift 1} &
\multicolumn{4}{c}{Shift 2} &
\multicolumn{4}{c}{Shift 3} \\
&
\multicolumn{1}{c}{time} &
\multicolumn{1}{c}{lag} &
\multicolumn{1}{c}{total} &
\multicolumn{1}{c}{actual} &
\multicolumn{1}{c}{time} &
\multicolumn{1}{c}{lag} &
\multicolumn{1}{c}{total} &
\multicolumn{1}{c}{actual} &
\multicolumn{1}{c}{time} &
\multicolumn{1}{c}{lag} &
\multicolumn{1}{c}{total} &
\multicolumn{1}{c}{actual} \\
\hline
MaMc & 1.63 & 4.1 & 5.7 & 5.4 & 1.05 &  5.1 &  6.1 &  5.7 & 4.44 & 25.2 & 28.5 & 28.8 \\
MaWz & 0.42 & 8.2 & 8.6 & 7.9 & 0.28 & 10.3 & 10.6 & 10.5 & 1.38 & 51.4 & 52.1 & 55.0 \\
McWz & 0.39 & 4.1 & 4.5 & 4.4 & 0.25 &  5.2 &  5.4 &  5.2 & 0.89 & 26.2 & 26.9 & 28.6 \\
\hline
	\end{tabular}
	\tablefoot{Percentage amplitude errors for 3 different shifts predicted due to (1) time-average smearing; (2) the triangular lag function of an FX correlator; (3) the product of (1) and (2).
	(4) The actual loss in amplitude observed averaged across all spectral points and time integrations.}
	\label{tab:shiftresultsamp}
\end{table*}

\Tab~\ref{tab:shiftresultsamp} compares the difference in amplitudes compared with those predicted in \eqns~\ref{eqn:timesmear} and \ref{eqn:lagcorr}.
Since the source is very strong, the large amplitude loss in the final shift still leaves a large SNR and so there is still good agreement between the phases in the two datasets.
\subsection{Searching for the three components}
\label{3comp}
\begin{figure}[t]
	\begin{center}
		\resizebox{\hsize}{!}{\includegraphics[trim=0 20 0 0, clip]{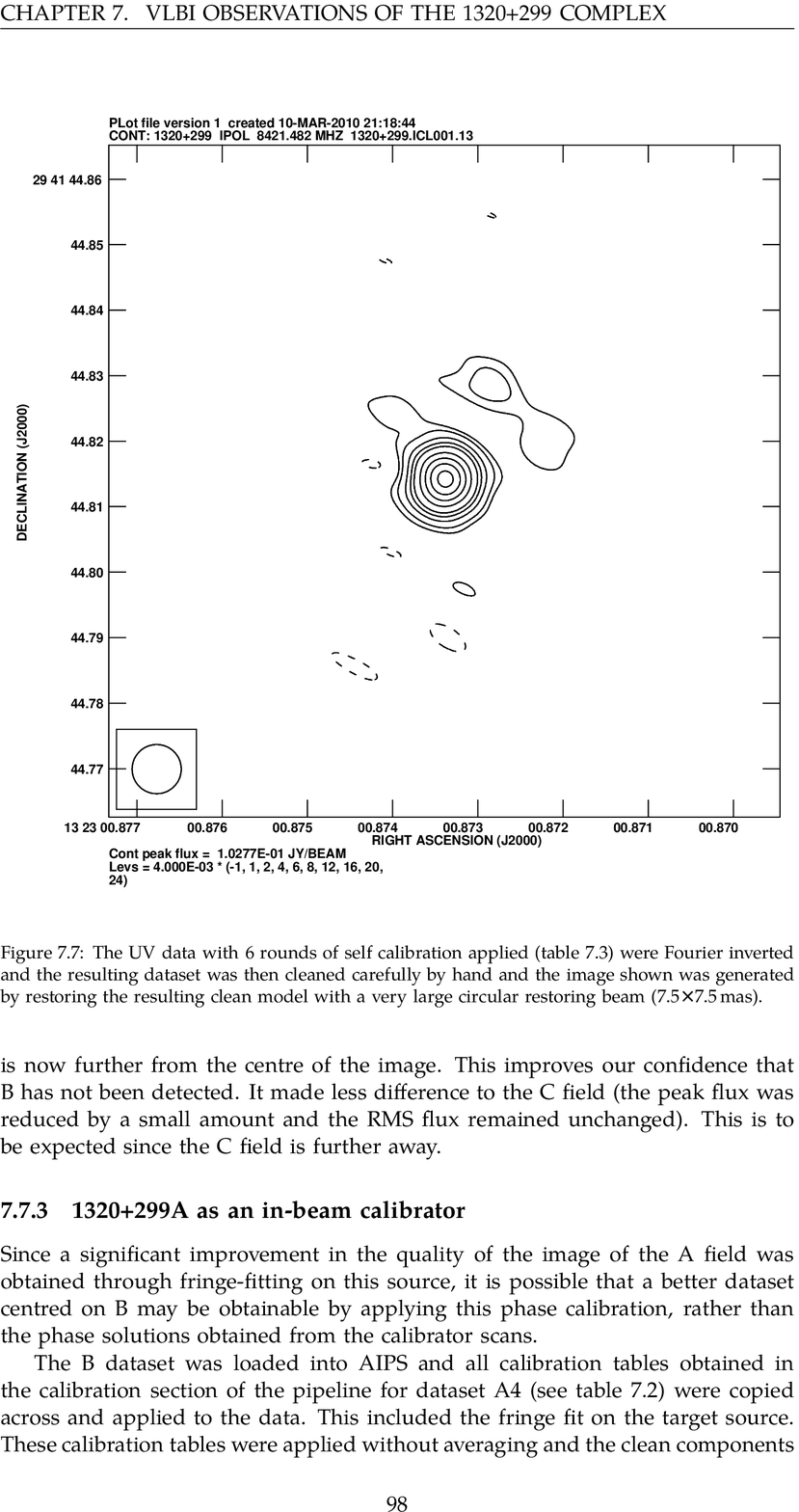}}
	\end{center}
	\caption{Map of 1320+299A centred on RA~13\rah23\ram0\fras8733 Decl.~+29\degr41\arcmin44\farcs814 after primary beam correction and self-calibration, imaged with a large circular restoring beam (7.5$\times$7.5\,mas); peak flux 102\,mJy beam$^{-1}$; contour levels -4, 4, 8, 16, 24, 32, 48, 64, 80, 96\,mJy beam$^{-1}$.}
	\label{fig:aselfcal}
\end{figure}

A full correlation of the target data was made with the phase centre close to the \apriori\ coordinates of B (the pointing centre).
An integration time of 0.2\,s and 512 channels per sub-band kept time-average and bandwidth smearing to an extremely low level.
The correction given in \eqn~(\ref{eqn:lagcorr}) was deemed unnecessary.
Three shifted datasets were then generated centred on the \apriori\ coordinates of the three components.
Imaging of the phase-reference calibrator showed it to be entirely unresolved.
The three shifted datasets were then calibrated in the standard way  using the phase reference source and then imaged.
Only component A was detected and after primary beam correction and self-calibration was revealed to be entirely unresolved (see \fig~\ref{fig:aselfcal}) with a flux density of 103\,mJy beam$^{-1}$ (VLA observations in Cornwell \etal\ imply a flux density of approximately 170\,mJy beam$^{-1}$).
No detection was made of the other two sources. 

\subsection{Imaging the entire primary beam}
\label{sec:allpb}
Finally, in order to search for any sources not present in previous studies over the relatively wide areas over which flux was recovered by Cornwell \etal and to test the feasibility of using UV shifting to generate extremely large images, the entire primary beam was imaged.
The `edge' of the primary beam for the array was arbitrarily chosen to be the point at which the Ef-Mc baseline had the same predicted sensitivity as the Wz-Ma baseline -- their relative sensitivities as a function of offset being calculated using the summation in \eqn~(\ref{eqn:noisearraywidefield}), assuming a radial sinc function for each antenna's voltage pattern. 
This leads to a radius of 90\arcsec\ from the pointing centre, approximately double the area enclosed by the HPBW of Effelsberg.

Shifting in turn to phase centres placed on a grid, and imaging each one allows the imaging of this entire field with each sub-image being produced from a small dataset.
Spacing the phase centres 5\arcsec\ apart in declination and 0.5\ras\ (equivalent to 6.5\arcsec) apart in right ascension gives a convenient grid and results in a total of 782 datasets (each 20MB in size) required to image the entire primary beam. 
With images of $4096 \times 4096$ pixels with each pixel having dimensions of $0.85 \times 0.7$ mas, the entire area is covered with a slight overlap between the images.
Averaging each sub-band to a single channel incurs bandwidth smearing of 4\% on this image.
Losses due to non-coplanar effects are $\sim$1\%.

In view of the strong detection of component A, two refinements were made to the calibration procedure:
(1) phase-reference solutions were derived from a fringe fit on component A\footnote{To ensure that the fringe-fit solutions corrected for unmodelled delays and \emph{not} any error in the \textit{a priori} position of component A, the astrometric position of component A, phase-referenced to J1329+3154 was carefully measured and another dataset was generated shifted to this precise position.};
(2) an attempt was made to subtract a simple model of component A from the UV datasets; this was found to reduce spurious bright sources significantly, particularly in the fields closest to component A.

The calibration and imaging of each sub-field was automated using ParselTongue \citep{Kettenis:2006}.
Since much of the calibration could be reused for each field, this process proceeded very quickly (11s per field).
Two normally-weighted images with no deconvolution were generated for each sub-field (one with and one without primary beam correction).
The only computationally demanding steps were generating the images. 

\begin{figure*}
	\centering
	\subfloat[]{\label{fig:peak}\includegraphics[width=6.5cm]{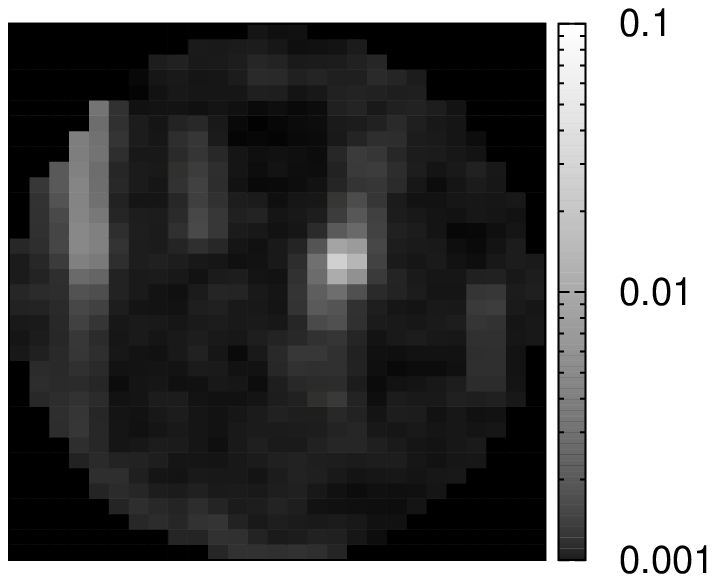}} \,
	\subfloat[]{\label{fig:rms}\includegraphics[width=6.5cm]{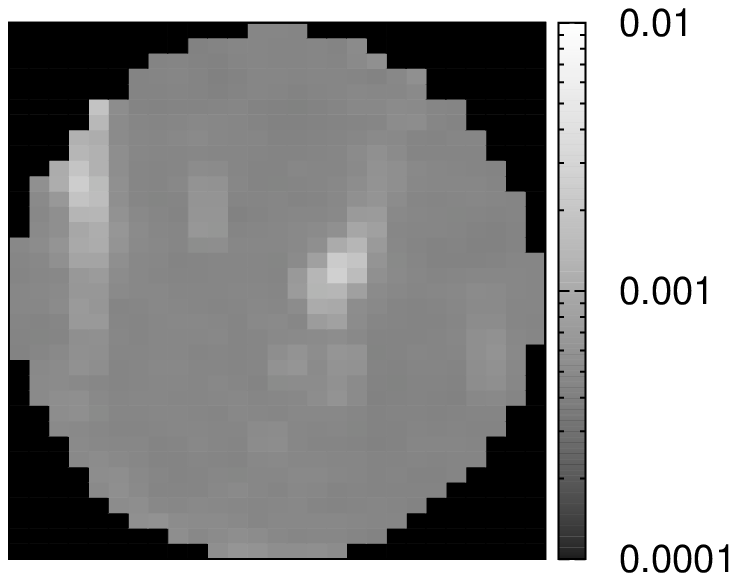}} \\
	\subfloat[]{\label{fig:peakpb}\includegraphics[width=6.5cm]{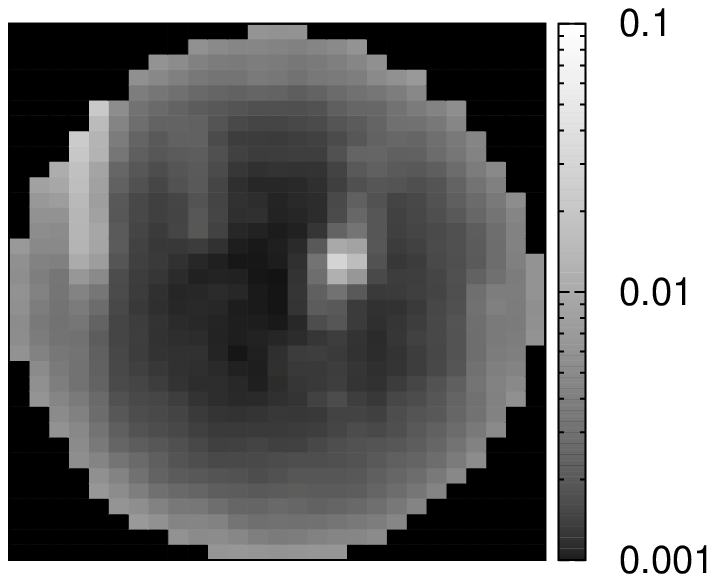}} \,
	\subfloat[]{\label{fig:rmspb}\includegraphics[width=6.5cm]{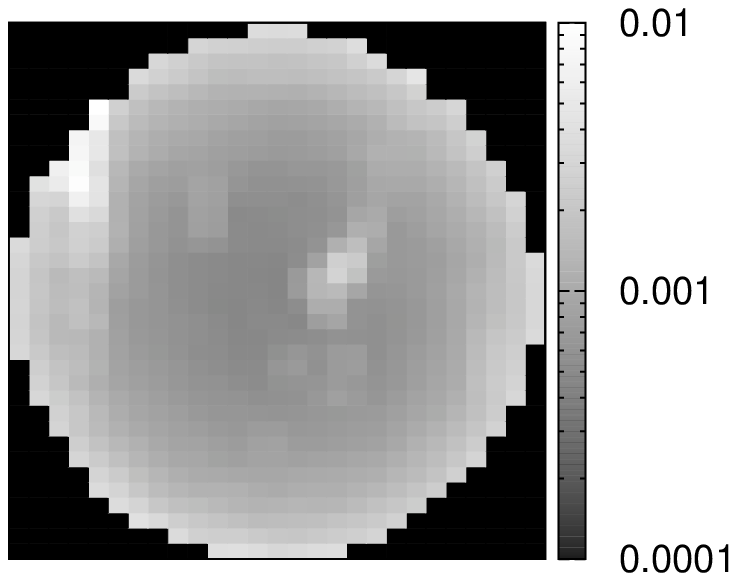}} \\
	\caption{Images representing the entire primary beam.
	Centred on RA 13\rah23\ram2\fras75 Decl. +29\degr41\arcmin32\farcs5.
	Each pixel represents a field 6.5\arcsec$\times$\,5\arcsec.
	The flux density in Jy beam$^{-1}$ is shown for the brightest pixel in each image and the RMS of all of the pixels both with and without primary beam correction applied: 
	(a) Peak flux density no PB correction; (b) RMS flux density no PB correction; (c) Peak flux density PB correction; (d) RMS flux density PB correction.
	The brightest point (towards the top-right) is the A component, attenuated since a simple model of A has been subtracted from the UV data.
	All flux apart from the brightest pixel in each image is spurious flux from the sidelobes of A or noise.}
	      \label{fig:corr1field}
\end{figure*}

\Fig~\ref{fig:corr1field} illustrates the results of this process.
Only a simple model of A has been subtracted so this source is clearly visible.
Other flux is very likely to be distant sidelobes of A.
Away from this spurious flux the RMS noise is approximately 500\,\textmu Jy beam$^{-1}$, approximately double the theoretical prediction and similar to that reached by Cornwell \etal.

\section{Discussion}
\subsection{The nature of the 1320+299 Complex}
Our observations aimed to clarify the nature of the 1320+299 complex since the detection (or non detection) of each component on milli-arcsecond scales would further aid their classification.

Component A was the only one detected by the present observations.
The image we obtained of component A (\fig~\ref{fig:aselfcal}) shows the core-jet structure usually found in flat spectrum radio quasars.
This finding confirms the interpretation suggested by Cornwell \etal\ considering their observations made at lower resolution, in which the A component also shows a core-jet structure with the core presenting a flat spectral index.

The lack of detection of any compact feature in component B may confirm the interpretation given by Cornwell \etal, who have suggested that it is an head-tail galaxy.
On the other hand, the non-detection of component C, suggested by Cornwell \etal\ to have an appearance similar to that of an edge-brightened hot-spot, is still puzzling.  

\subsection{Accurate UV shifting}
We have outlined a method whereby a technically inclined astronomer can image any part of the primary beam of interest. 
Both the accuracy of this method (and the ability of the DiFX software correlator to generate visibility datasets with sufficient resolution) have been demonstrated for shifts significantly larger than those that could be required for Earth-based VLBI with parabolic dishes.
With the exception of unavoidable instrumental effects (the primary beam) the data are not degraded in any significant way compared with carrying out separate observations.
This drastically increases the number of objects which can be studied and allows VLBI surveys to be carried out.
No correlation facilities or expertise are required but substantial storage and computing resources would be required for larger datasets.

\subsection{Future Work}
The main problem with this approach as it stands is that the astronomer still requires access to a sophisticated delay model. 
Code for generating these models, while necessary for correlation, is not particularly portable nor easy to assimilate into existing data analysis software.

This algorithm has been implemented with full accuracy within the DiFX correlator itself \citep{Deller:2011}.
For a large subset of \widefield\ VLBI observations this simplifies the process greatly, obviating the need for large datasets to be written to disk at all, and providing the astronomer with a large number of standard VLBI datasets.
This is particularly appropriate for cases where the field of interest has already been studied at low resolution with at least equivalent sensitivity to the VLBI observation.
VLBI surveys of objects detected with other instruments can provide very useful complementary informations for mult-wavelength studies, for example unambiguously identifying AGN \citep{Middelberg:2010}.

In other cases there will be a benefit to imaging the entire primary beam.
With the increasing sensitivity of VLBI, it may often be the case that pre-existing lower-resolution images are not as sensitive as the VLBI studies, at least until instruments such as the EVLA and e-MERLIN have completed their legacy surveys.
Transient phenomena (such as supernovae) may also be detected in this way, either with dedicated observations or by commensally reprocessing other observations using \widefield\ techniques.
Finally, when extended objects are imaged at low resolution, large swathes of sky may have to be imaged to find compact emission (for example searching for hotspots in jets or supernovae and supernova remnants close to a galactic centre.

For cases where the shifting must be done after correlation, it is clear that existing shifting algorithms are inadequate.
\begin{figure}
\resizebox{\hsize}{!}{\includegraphics[clip]{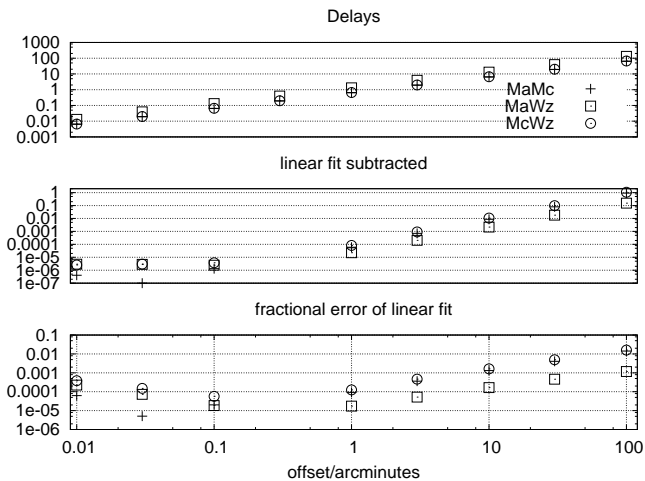}}
\caption{An illustration of the non-linearity of the delay (in micro\-seconds) with offset for three baselines. Matera-Medicina (crosses), Matera-Wettzell (squares), Medicina-Wettzell (circles). The top panel shows the delay at offsets between 0.01 and 100 arcminutes compared to the delay at zero offset. The middle panel shows the same delays after the subtraction of the linear interpolation between zero offset and the 0.3 arcminute value. The bottom panel shows the quotient of the first two plots i.e. the fractional error of the linear fit.}
\label{fig:delays}
\end{figure}
\Fig~\ref{fig:delays} shows the limitation of using a linear delay model (note that the linear fit calculated here is identical to that used to calculate the $u$ and $v$ coordinates in DiFX).
A linear relationship between delay and offset (implicitly assumed in interferometry imaging) results in an positional error approaching 1\% of the shift within the primary beam for a typical VLBI observation.

As explained in \sect~\ref{derivation}, current VLBI datasets contain the delay ($w$) and the partial derivatives of delay with respect to $l$ and $m$ ($u$ and $v$).
By adding further derivatives or polynomial functions of the delay with respect to $l$ and $m$, the delay across the wide field could be fully characterised and accurate UV shifting to any point in the wide field could be done at any stage during correlation, calibration or imaging.

At the calibration stage, this three-dimensional ($l$, $m$ and time) model could be refined using in-beam calibrators and/or observing techniques which use observations of multiple calibrators.
This technique could also be useful for low-frequency interferometry where the unmodelled delay is known to vary significantly and non-linearly across the primary beam.

Calibration using multiple weak sources across the primary beam is described in \citet{Garrett:2004}
We have had some early success applying a variation of this technique to other wide field datasets, and it is likely to become ever more useful as VLBI bandwidths increase.
This opens up the possiblity of obtaining noise-limited images and high-accuracy astrometry from fields where there is no nearby calibrator.

In conclusion, advances in correlation and data reduction techniques has made VLBI surveys possible and even routine, without any loss in sensitivity (or astrometric accuracy) apart from that imposed by the instrument.
Imaging any or all points within the interferometer elements' primary beam with VLBI resolution will undoubtedly provide a useful complement to wide field surveys at other wavelengths.
In addition, data reduction techniques developed for wide field VLBI data volumes will no doubt be applicable to new wide field and wide band radio telescopes.

\bibliographystyle{hapj} 
\bibliography{mn-jour,tech}
\begin{acknowledgements}
John Morgan's research was supported by the EU Framework 6 Marie Curie Early Stage Training programme under contract number MEST-CT-2005-19669 `ESTRELA'.
This work is based on observations made with the Medicina telescope operated by INAF-Istituto di Radioastronomia, the 100\,m telescope of the MPIfR (Max-Planck-Institut f\"ur Radioastronomie) at Effelsberg, the VLBI antenna at the geodetic observatory in Wettzell and the VLBI antenna at the Centre of Spatial Geodesy, Matera.
We would like to thank the directors and staff of the participating observatories for their kind collaboration in this project.  
John Morgan would like to thank Richard Porcas for useful discussions on the definition of time in VLBI observations.
\end{acknowledgements}
\end{document}